\documentclass[%
reprint,
amsmath,amssymb,
aps,
]{revtex4-1}

\usepackage{graphicx}% Include figure files
\usepackage{dcolumn}% Align table columns on decimal point
\usepackage{bm}% bold math
\usepackage{mathrsfs}
\begin{document}
	
	\title{\large\bf Collapse of the Fano Resonance Caused by the Nonlocality of the Majorana State}
	
	\author{S.\,V.\, Aksenov$^{1}$}%
	\email{asv86@iph.krasn.ru}
	\author{M.\,Yu.\, Kagan$^{2,3}$}
	\email{kagan@kapitza.ras.ru}
	
	\affiliation{%
		$^1$ Kirensky Institute of Physics, Federal Research Center KSC SB RAS, 660036 Krasnoyarsk, Russia\\
		$^2$ P.L. Kapitza Institute for Physical Problems RAS, 119334 Moscow, Russia\\
		$^3$ National Research University Higher School of Economics, 101000 Moscow, Russia
	}
	
%	\date{\today}% It is always \today, today,
	
	\begin{abstract}
		
One of the main features of the Majorana state, which attracts a considerable current interest to these excitations in solid-state systems, is related to its nonlocal character. It is demonstrated that the direct consequence of such nonlocality is the collapse of the Fano resonance manifesting itself in the conductance of an asymmetric interference device, the arms of which are connected by a one-dimensional topological superconductor. In the framework of the spinless model, it is shown that the predicted effect is associated with an increase in the multiplicity of the degeneracy of the zero-energy state of the structure arising at a specific case of the Kitaev model. Such an increase leads to the formation of a bound state in the continuum.
		
		\begin{description}
			\item[PACS number(s)]
			71.10.Pm, % Fermions in reduced dimensions (anyons, composite fermions, Luttinger liquid, etc.)
			74.78.Na, % Mesoscopic and nanoscale systems
		\end{description}
	\end{abstract}

\maketitle

%\tableofcontents

\maketitle

%\tableofcontents
{\bf 1.} The formation of bound states in a continuum (BSC) taking place in quantum systems is a particular case of the coupling between the states of continuous and discrete spectra, when their hybridization vanishes \cite{neumann-29,hsu-16}. The BSC can arise due to both fundamental mechanisms related to a certain symmetry of the structure \cite{schult-89} and to an accidental vanishing of the mentioned coupling occurring in the course of continuous change in the parameters of the system \cite{friedrich-85}. In the ideal case, the systems with BSC should exhibit infinite Q factors, which makes them quite promising for the optical applications, such as lasers, filters, and detectors \cite{foley-14}.

Quantum dot arrays are popular objects often exhibiting BSC \cite{kagan-18}. This is already evident in the simplest case of two dots whose eigenstates can be considered as bonding and antibonding ones. Then, breaking the symmetry of the open system by the continuous variation of the parameters characterizing the tunnel coupling of the double quantum dot with contacts, we can trace the crossover from the situation, when the antibonding state is a BSC at the symmetrical parallel connection, to that, when both these states have the same finite lifetime at the serial connection \cite{guevara-03}.

In the intermediate case of an asymmetric parallel connection, the conductance resonance related to the antibonding state has the form of a Fano resonance \cite{fano-61}, the width of which is directly proportional to the value of hybridization of this state with the continuum. A similar picture is observed when the Aharonov-Bohm phase is taken into account \cite{orellana-04,lu-05}. As a result, in highly asymmetric transport geometry, the conductance is characterized by the presence of a wide Breit-Wigner resonance and a narrow Fano resonance, similar to the Dicke effect in optics \cite{dicke-53}. Thus, the Fano resonance can be interpreted as some precursor of BSC, and its collapse is a point in the parametric space, where the BSC appears \cite{kim-99a,kim-99b}. An increase in the number of quantum dots in the structure leads to the corresponding increase in the number of BSCs \cite{gong-09}. The inclusion of many-particle effects also leads to the formation of additional BSCs and Fano resonances \cite{sadreev-08,kagan-17a}. In their turn, the spin-orbit coupling and Zeeman splitting make it possible to implement the spin filtering effect based on these features \cite{vallejo-10,kagan-17b}. Note that BSCs arise in the systems under discussion naturally because these systems are not one-dimensional in the coordinate or energy space \cite{lee-99,sadreev-03}.

The phenomenon of topological superconductivity attracts a considerable current interest since it is promising mainly for quantum computations, which are resistant to the processes disturbing the phase of a qubit state. One of the scenarios allowing the actual formation of the Majorana state (MS) in 1D systems is the combination of three factors: spin-orbit coupling, superconducting pairing, and magnetic field \cite{lutchyn-10,oreg-10,valkov-19b}. In this case, at a certain relation between the parameters in the normal phase, an odd number of Fermi points arise at $k\geq0$ ($k$ is the wave vector). As a result, the superconducting pairing of electrons belonging to one subband takes place, i.e., we obtain the effective $p$-wave pairing. Thus, the wire becomes equivalent to the Kitaev chain - an idealized 1D system, for which the formation of MS was demonstrated for the first time \cite{kitaev-01}. The importance of spin-orbit interaction is also emphasized by the fact that the self-conjugated operator of the quasiparticle excitation with zero energy cannot have the form $\beta=ua_{\uparrow}+va_{\downarrow}^{+}$. Several experiments on the tunneling spectroscopy at InAs and InSb semiconducting wires with the strong spin-orbit coupling and induced superconducting pairing (further on, for brevity, we will use the abbreviation superconducting wire, SW) provide an evidence in favor of the MS formation in the aforementioned structures \cite{mourik-12}. 

\begin{figure}[tb]
	\includegraphics[width=0.45\textwidth]{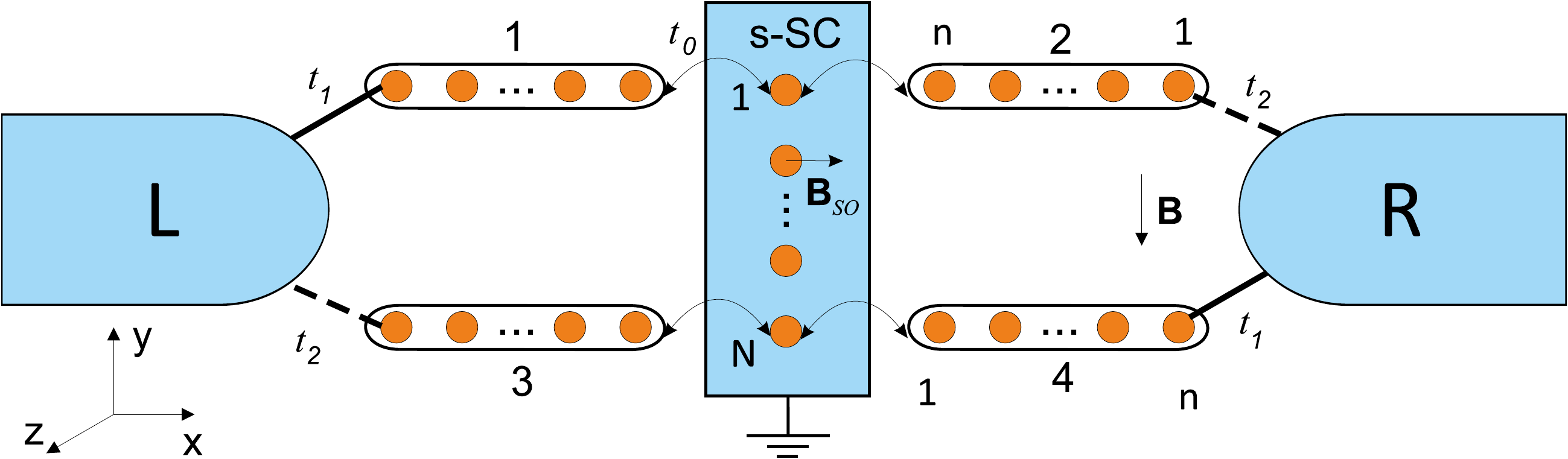}
	\caption{\label{model} Fig. \ref{model}. Aharonov-Bohm ring with the arms connected by the superconducting wire exhibiting the Rashba spin-orbit coupling.}
\end{figure}
However, despite the progress in epitaxial growth techniques and measurement methods \cite{zhang-18}, the obtained quantization of conductance at zero voltage does not provide a sufficient evidence for the formation of a topologically nontrivial phase in the structures under discussion \cite{liu-17}. As a result, today the search for alternative ways of detecting MS \cite{valkov-17}, in particular, using the nonlocal nature of this excitation is a rather topical task. In \cite{valkov-19}, the transport characteristics of the symmetric Aharonov-Bohm ring, the arms of which are connected by a SW bridge (see Fig. \ref{model} at $t_{1}=t_{2}$), are analyzed. It is shown that if the nontrivial phase arises in the wire, then the conductance exhibits the Dicke effect. Moreover, the properties of the Fano resonance depend on the overlap of the Majorana wave functions localized at opposite ends of the SW.

In our work, we study the asymmetric ring shown in Fig. \ref{model}, where asymmetry corresponds to unequal values of the tunneling parameters between the contacts and the device, $t_{1}\neq t_{2}$. It is shown that, in contrast to the previously studied symmetric geometry \cite{valkov-19}, new Fano resonances appear in the conductance of an asymmetric device. It is found that their width is directly proportional to the degree of nonlocality of the state of SW with the lowest energy. In other words, the higher is the near-edge probability density, the narrower is the Fano resonance. As a result, in the limiting case of two noninteracting Majorana fermions, this conductance feature disappears.

{\bf 2.} The quantum transport features discussed below are related to the involvement of SW. The SW Hamiltonian has the form
\begin{eqnarray} \label{HW}
&&\hat{H}_{W} =\sum\limits_{j=1}^{N}\left[\sum\limits_{\sigma}\xi a^+_{j\sigma}a_{j\sigma}+
\left(\Delta a_{j\uparrow}^{+}a_{j\downarrow}^{+} + ih a^+_{j\uparrow}a_{j\downarrow} + h.c.\right)\right]+\nonumber\\
&&~~~~~~~~~~~~+\frac{1}{2}\sum\limits_{\sigma;j=1}^{N-1}\left[-ta^+_{j\sigma}a_{j+1,\sigma}+i\alpha\sigma a^+_{j\overline{\sigma}}a_{j+1,\sigma}+h.c.\right],~~
\end{eqnarray}
where $\xi=\epsilon_{d}-\mu$ is the on-site energy controlled by the gate electric field, characterized by energy, $\epsilon_{d}$; $\mu$ is the chemical potential; $t$ is the hopping integral for the nearest sites; $\alpha$ is the magnitude of the Rashba spin?orbit coupling; $\Delta$ is the parameter characterizing the s-wave superconducting pairing; $h$ is the Zeeman energy related to the magnetic field $\textbf{B}$ in the device plane. Then, the topologically nontrivial phase takes place if the following inequality is valid \cite{lutchyn-10,oreg-10}
\begin{equation} \label{TPineq}
\left(\xi-t\right)^{2}+\Delta^{2}<h^{2}<\left(\xi+t\right)^{2}+\Delta^{2}.
\end{equation}
Note that although $\alpha$ is formally not involved in \eqref{TPineq}, a nonzero spin?orbit coupling is essential for the formation of MS, as we have already mentioned in Section 1. Moreover, the effective Rashba field, $\textbf{B}_{SO}$, should be oriented along the perpendicular to the Zeeman field $\textbf{B}$. Further on, in our calculations, all energy parameters are measured in the units of $t$: $t=1$, $\Delta=0.25$, $\alpha=0.2$, $\mu=0$.

The wires in the normal phase (NW), which are the arms of the ring (see Fig. \ref{model}), are assumed to be the same. Their Hamiltonians, $\hat{H}_{1-4}$, are obtained from \eqref{HW} at $\Delta=\alpha=0$. The coupling between the SW and NWs is described by the tunneling Hamiltonian,
\begin{eqnarray} \label{HWl}
&&\hat{H}_{T} =-t_{0}\sum\limits_{\sigma}\left[\left(b_{Ln\sigma}^{+}+b_{Rn\sigma}^{+}\right)a_{1\sigma}
\right. + \\
&&~~~~~~~~~~~~~~~~~~~~~~~~~~~+\left.\left(d_{L1\sigma}^{+}+d_{R1\sigma}^{+}\right)a_{N\sigma}\right]+ h.c.,\nonumber
\end{eqnarray}
where $t_{0}$ is the hopping integral between the edge SW and NW sites; $b_{L\left(R\right)n\sigma}^{+}$ is the creation operator for an electron with spin projection $\sigma$ at the last site in the left (right) upper NW; $d_{L\left(R\right)1\sigma}^{+}$ is the creation operator for an electron with spin projection $\sigma$ at the first site in the left (right) lower NW. In its turn, the coupling between the device (SW$+$NW) and contacts is also described by the tunneling Hamiltonian, which at the same time plays the role of interaction operator when we use the diagram technique for nonequilibrium Green's functions
\begin{eqnarray} \label{V}
&&\hat{V} =-\sum \limits_{k\sigma}\left[c_{Lk\sigma}^{+}\left(t_{1}b_{L1\sigma}+
t_{2}d_{Ln\sigma}\right)\right. + \nonumber\\
&&~~~~~~~~~~~~~~~~~+\left. c_{Rk\sigma}^{+}\left(t_{2}b_{R1\sigma} +
t_{1}d_{Rn\sigma}\right)\right]+ h.c.,
\end{eqnarray}
where $c_{L\left(R\right)k\sigma}^{+}$ is the creation operator for an electron with wave vector $k$ and spin projection $\sigma$ at the left (right) contact; $t_{1,2}$ are the hopping integrals between the contacts and device. Hamiltonian for $i$th contact ($i=L,~R$) has the simple form $\hat{H}_{i}=\sum_{k}\left(\epsilon_{k}-\mu_{i}\right)c_{ik\sigma}^{+}c_{ik\sigma}$,where $\mu_{L,R}=\mu \pm eV/2$ is the electrochemical potential of the contact including the applied bias voltage.

To calculate the steady-state current flowing across the device, it is convenient to diagonalize its Hamiltonian, $\hat{H}_{D}=\hat{H}_{W}+\sum\limits_{i=1}^{4}\hat{H}_{i}+\hat{H}_{T}$, using the Nambu operators in the site representation, $\hat{f}_{j}=\left(f_{j\uparrow}~f_{j\downarrow}^{+}~ f_{j\downarrow}~f_{j\uparrow}^{+}\right)^T$, where $f_{j\sigma}$ is the annihilation operator of an electron with spin projection $\sigma$ at jth site of NW or SW \cite{valkov-19}. Then, we can specify the matrix Green's function of the ring in the following form
\begin{equation} \label{GF}
\hat{G}^{ab}\left(\tau,\tau'\right)=-i\left\langle T_{C}\hat{\Psi}\left(\tau_{a}\right)\otimes
\hat{\Psi}^{+}\left(\tau'_{b}\right)\right\rangle,~
\end{equation}
where $T_{C}$ is the ordering operator at the Keldysh time contour consisting of the lower (superscript $+$) and upper (superscript $-$) parts  \cite{keldysh-64}; $a,b=+,-$; $\hat{\Psi}$ has the dimension $4\left(N+4n\right) \times 1$, i.e. it includes the Nambu operators for both SW and all NWs,
\begin{equation} \label{Psi}
\hat{\Psi}=\left(\hat{b}_{L1}...\hat{b}_{Ln}\hat{d}_{L1}...\hat{d}_{Ln}\hat{a}_{1}...\hat{a}_{N}\hat{b}_{R1}...\hat{b}_{Rn}\hat{d}_{R1}...\hat{d}_{Rn}\right)^{T}.
\end{equation}

The electron current in the left contact is written as  $I=e\left\langle\dot{N}_{L}\right\rangle$ ($N_{L}=\sum_{k\sigma}c_{Lk\sigma}^{+}c_{Lk\sigma}$ is the particle number operator in the left contact). The solution of the Heisenberg equation gives ($\hbar=1$)
\begin{eqnarray} \label{IL1}
&&I=
2e\sum\limits_{k}Tr\Biggl[\hat{\sigma}Re\Biggl\{\hat{t}_{1}^{+}\left(t\right)\hat{G}_{k,L1}^{+-}\left(t,t\right)+\Biggr.\Biggr.\\
&&\Biggl.\Biggl.~~~~~~~~~~~~~~~~~~~~~~~~~~~~~~~~~~~~+\hat{t}_{n}^{+}\left(t\right)\hat{G}_{k,Ln}^{+-}\left(t,t\right) \Biggr\} \Biggr],\nonumber
\end{eqnarray}
where $\hat{\sigma}=diag\left(1,-1,1,-1\right)$; diagonal matrices $\hat{t}_{1,n}$ are the functions of time resulting from the unitary transformation \cite{rogovin-74} converting the voltage dependence into the operator the operator $\hat{V}$,
\begin{equation} \label{t12}
\hat{t}_{1,n}=\frac{t_{1,2}}{2}diag\left(e^{-i\frac{eVt}{2}},e^{i\frac{eVt}{2}},e^{-i\frac{eVt}{2}},e^{i\frac{eVt}{2}}\right)\cdot\hat{\sigma}
\end{equation}
In \eqref{IL1}, the mixed Green's functions have the form  $\hat{G}_{k,L1}^{+-}=i\left\langle \hat{b}_{L1}^{+}\otimes\hat{c}_{Lk}\right\rangle$ and $\hat{G}_{k,Ln}^{+-}=i\left\langle \hat{d}_{Ln}^{+}\otimes\hat{c}_{Lk}\right\rangle$. In the Nambu operator space, $\hat{H}_{D}$ has the form of Hamiltonian for free particles, therefore, in specifying averages in $\hat{G}_{k,L1}^{+-}$ and $\hat{G}_{k,Ln}^{+-}$ we should use the same guidelines as those for the averages for $T_{C}$-ordered product of the second quantization operators \cite{vonsovsky-77,arseev-15}. As a result, at $t\rightarrow0$ expression \eqref{IL1} transforms to
\begin{eqnarray} \label{IL2}
&&I=
2e\int\limits_{C}d\tau_{1}Tr\Biggl[\hat{\sigma}Re\Biggl\{\hat{\Sigma}_{L1,L1}^{+a}\left(-\tau_{1}\right)\hat{G}_{L1,L1}^{a-}\left(\tau_{1}\right)+\Biggr.\Biggr.\\
&&\Biggl.\Biggl.+\hat{\Sigma}_{Ln,Ln}^{+a}\left(-\tau_{1}\right)\hat{G}_{Ln,Ln}^{a-}\left(\tau_{1}\right)+\hat{\Sigma}_{L1,Ln}^{+a}\left(-\tau_{1}\right)\hat{G}_{Ln,L1}^{a-}\left(\tau_{1}\right)+\Biggr.\Biggr.\nonumber\\
&&~~~~~~~~~~~~~~~~~~~~~~~~~~~~~~~~~~\Biggl.\Biggl.+\hat{\Sigma}_{Ln,L1}^{+a}\left(-\tau_{1}\right)\hat{G}_{L1,Ln}^{a-}\left(\tau_{1}\right) \Biggr\} \Biggr],\nonumber
\end{eqnarray}
where $\hat{\Sigma}_{Li,Lj}^{+a}\left(-\tau_{1}\right)=\hat{t}_{i}^{+}\left(0\right)\hat{g}_{Lk}^{+a}\left(-\tau_{1}\right)\hat{t}_{j}\left(\tau_{1}\right)$ are self-energies of the left contact ($i,j=1,n$); $\hat{g}_{Lk}^{+a}\left(-\tau_{1}\right)$ is the bare Green's function of the left contact. Integrating over time $\tau_{1}$ and using the Fourier transform, we get
\begin{eqnarray} \label{IL3}
&&I=
e\sum_{i,j=1,n}\int\limits_{-\infty}^{+\infty}\frac{d\omega}{\pi}
Tr\Biggl[\hat{\sigma}Re\Biggl\{
\hat{\Sigma}_{Li,Lj}^{r}\left(\omega\right)\hat{G}_{Lj,Li}^{+-}\left(\omega\right)+ \Biggr.\Biggr.\nonumber\\
&&\Biggl.\Biggl.~~~~~~~~~~~~~~~~~~~~~~~~~~~~~~~~
+\hat{\Sigma}_{Li,Lj}^{+-}\left(\omega\right)
\hat{G}_{Lj,Li}^{a}\left(\omega\right) \Biggr\}\Biggr].
\end{eqnarray}
The further transformation of \eqref{IL3} makes it possible to obtain an explicit form of the components associated with the local Andreev reflection and the nonlocal transfer of charge carriers. However, these expressions are quite cumbersome, and we do not present them here.

Note that many-particle interactions are absent in the system, therefore the Green's functions in the integrand \eqref{IL3} are determined taking into account all the tunneling processes between the device and contacts \cite{arseev-15}. In particular, $\hat{G}_{Lj,Li}^{a}$ block-matrices of the advanced Green's function of the whole device, $\hat{G}^{a}$, are determined by the Dyson equation,
\begin{equation}\label{Ga}
\hat{G}^{a}=\left[\left(\omega-\hat{h}_{D}-\hat{\Sigma}^{r}\left(\omega\right)\right)^{-1}\right]^{+},
\end{equation}
where $\hat{\Sigma}^{r}\left(\omega\right)$ is the matrix of retarded self-energy revealing the effect of both contacts on the ring. In the course of further numerical calculations, we will use the popular approximation of wide-band contacts, for which the real parts of the self-energy functions can be neglected and the imaginary parts can be considered as constants (see, for example, \cite{arseev-12}). Then, we have the following nonzero blocks $\hat{\Sigma}^{r}$:
\begin{eqnarray}\label{Sr}
&&\hat{\Sigma}^{r}_{L1,L1}=\hat{\Sigma}^{r}_{Rn,Rn}=-\frac{i}{2}\hat{\Gamma}_{11}, \hat{\Sigma}^{r}_{R1,R1}=\hat{\Sigma}^{r}_{Ln,Ln}=-\frac{i}{2}\hat{\Gamma}_{22},\nonumber\\
&&\hat{\Sigma}^{r}_{L1,Ln}=\hat{\Sigma}^{r}_{R1,Rn}=\hat{\Sigma}^{r}_{Ln,L1}=\hat{\Sigma}^{r}_{Rn,R1}=-\frac{i}{2}\hat{\Gamma}_{12},
\end{eqnarray}
where $\hat{\Gamma}_{ii}=\Gamma_{ii}\hat{I}_{4}$, $\Gamma_{ii}=2\pi t_{i}^{2}\rho$ is the function characterizing the broadening of energy levels of the device due to its interaction with the contact ($i=1,2$); $\rho$ is the density of states in the contact; $\Gamma_{12}=\sqrt{\Gamma_{11}\Gamma_{22}}$; $\hat{I}_{4}$ is the unit $4\times4$ matrix. Considering directly the asymmetric (symmetric) ring, we assume that $\Gamma_{22}=\Gamma_{11}/2=0.01$ ($\Gamma_{22}=\Gamma_{11}=0.01$).

The $\hat{G}_{Li,Lj}^{+-}$ blocks in \eqref{IL3} are obtained by the solution of Keldysh equation, $\hat{G}^{+-}=\hat{G}^{r}\hat{\Sigma}^{+-}\hat{G}^{a}$. Note that we consider the regime, when all the transient processes have ended and the bare Green's functions of the device are not involved to this equation \cite{arseev-15}. Here, nonzero blocks $\hat{\Sigma}^{+-}$ are given by the expressions
\begin{eqnarray}
&&\hat{\Sigma}_{\alpha i,\alpha j}^{+-}=-2\hat{\Sigma}_{\alpha i,\alpha j}^{r}
\hat{F}_{\alpha},~~~~\alpha=L,R,~~~~i,j=1,n,\nonumber\\
&&\hat{F}_{L\left(R\right)}=diag\biggl(n\left(\omega \pm eV/2\right),~n\left(\omega \mp eV/2\right),\biggr.\\
&&\biggl.~~~~~~~~~~~~~~~~~~~~~~~~~~~~~~~~~~n\left(\omega \pm eV/2\right),~n\left(\omega \mp eV/2\right)\biggr),\nonumber
\end{eqnarray}
where $n\left(\omega \pm eV/2\right)$ are the Fermi-Dirac functions.

{\bf 3.}
\begin{figure}[htbp]
	\includegraphics[width=0.525\textwidth]{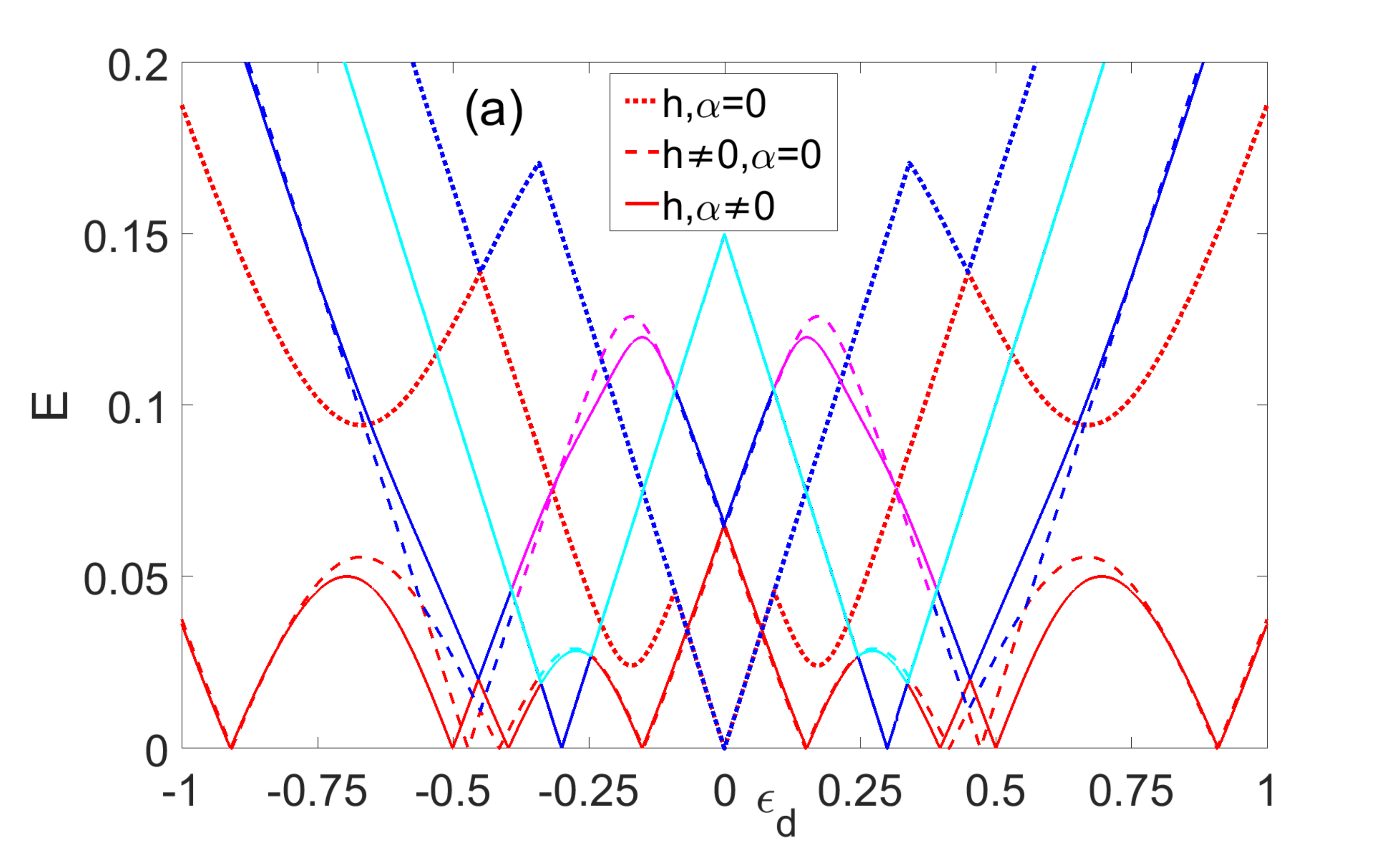}
	\includegraphics[width=0.525\textwidth]{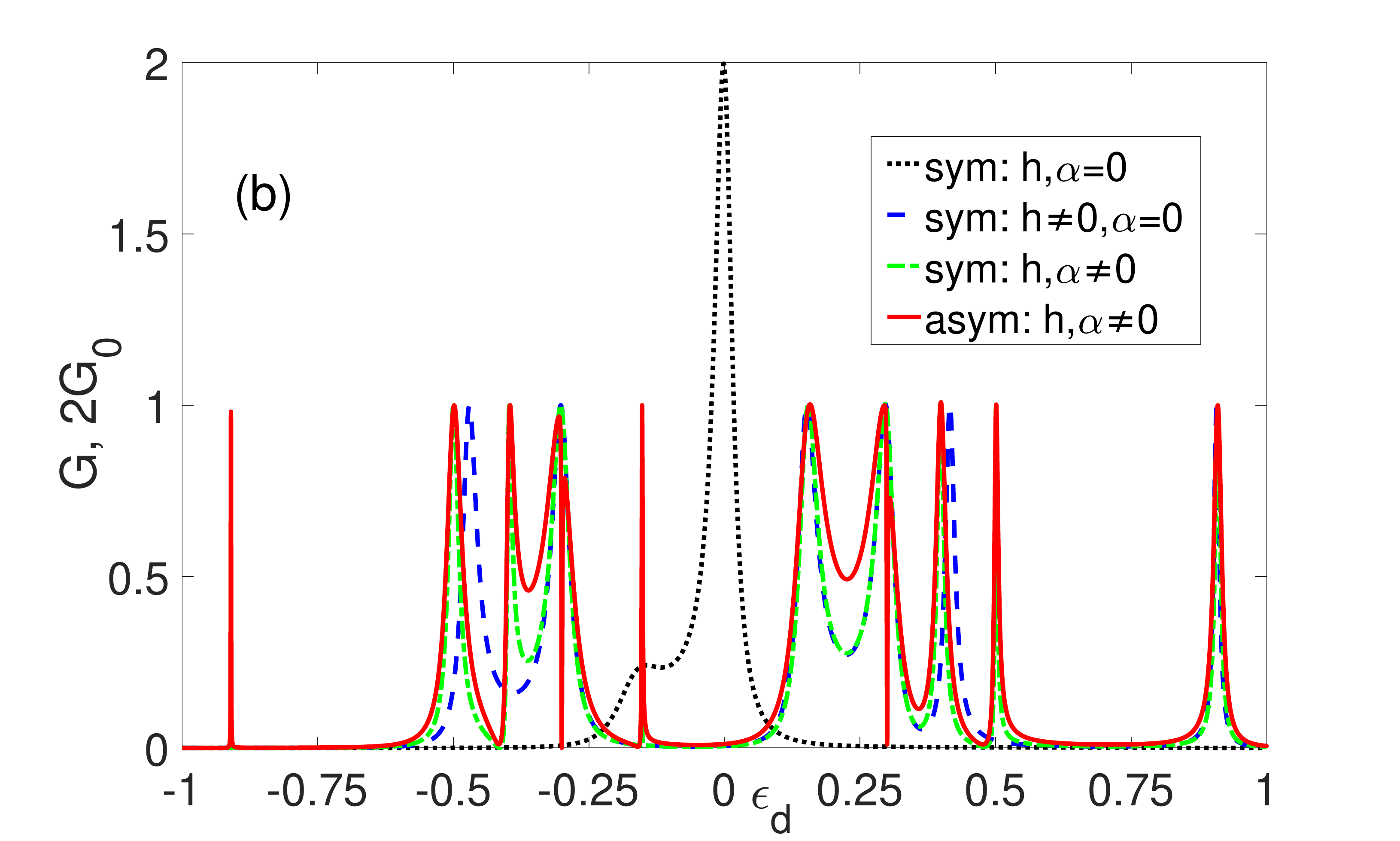}
	\caption{\label{6QD} Fig. \ref{6QD}. Excitation energies, $E_{1-4}$, ((a) and conductance (b) of the ring consisting of six sites versus electric field energy in the gate. Parameters: $n=1$, $N=2$, $t_{0}=0.5$, $h=0.3$.}
\end{figure}
Let us now turn to the results of the numerical calculation of quantum transport in the linear response regime and low temperatures ($eV,~k_{B}T\approx0$) for the system shown in Fig. \ref{model}. First, we consider the limiting case of a ring with a minimum number of sites, $n=1$, $N=2$. In Fig. \ref{6QD}a, we show the dependence of the energies of the first four states, $E_{1-4}$, on the gate field energy. At $h,~\alpha=0$ the energies are pairwise degenerate (see thin dotted curves). In addition, since the superconducting pairing in the ring is inhomogeneous, the gap arises at $\epsilon_{d}\neq0$. However, at zero gate field, we have $E_{1-4}=0$ and hence, the conductance,$G=dI/dV$, exhibits resonance only at $\epsilon_{d}=0$ (see the dotted line in Fig. \ref{6QD}b).

At $h>\Delta$, the gap is suppressed, and also the number of zeros in the spectrum becomes doubled due to the Zeeman splitting (see the dashed lines in Fig. \ref{6QD}a). As a result, the number of conductance peaks increases; this is illustrated by the dashed curve in Fig. \ref{6QD}b. However, not all zeros in the excitation energies manifest themselves as resonances in the conductance, which is a signature of arising BSC \cite{guevara-03,lu-05}. There are several ways to achieve a finite value of their lifetime. For example, it is possible to break the spatial symmetry of the eigenstates in the ring by introducing the spin-orbit coupling \cite{nowak-11}. As a result, the zeros of the excitation spectrum associated with the SW undergo a small shift, and Fano resonances arise in the conductance (see the solid and dash-dot curves in Figs. \ref{6QD}a and \ref{6QD}b, respectively). Thus, to have the Dicke effect in a symmetric ring with the superconducting central region, we need the combined effect of the magnetic field and spin-orbit coupling.

Note that at $\epsilon_{d}=\pm h$ the zero-energy state remains double degenerate even at $\alpha\neq0$. Such degeneracy is also related to the symmetry of the ring under study and implies the existence of additional BSCs \cite{volya-03,sadreev-06}. Their existence can manifest itself in the conductance, if we introduce an asymmetry of the tunneling parameters to the contacts. Solid curve in Fig. \ref{6QD}b shows that in this case, additional Fano resonances appear at $\epsilon_{d}=\pm h$. A similar effect arises if the Aharonov-Bohm phase is taken into account \cite{orellana-04,lu-05}.

\begin{figure}[htbp]
	\includegraphics[width=0.5\textwidth]{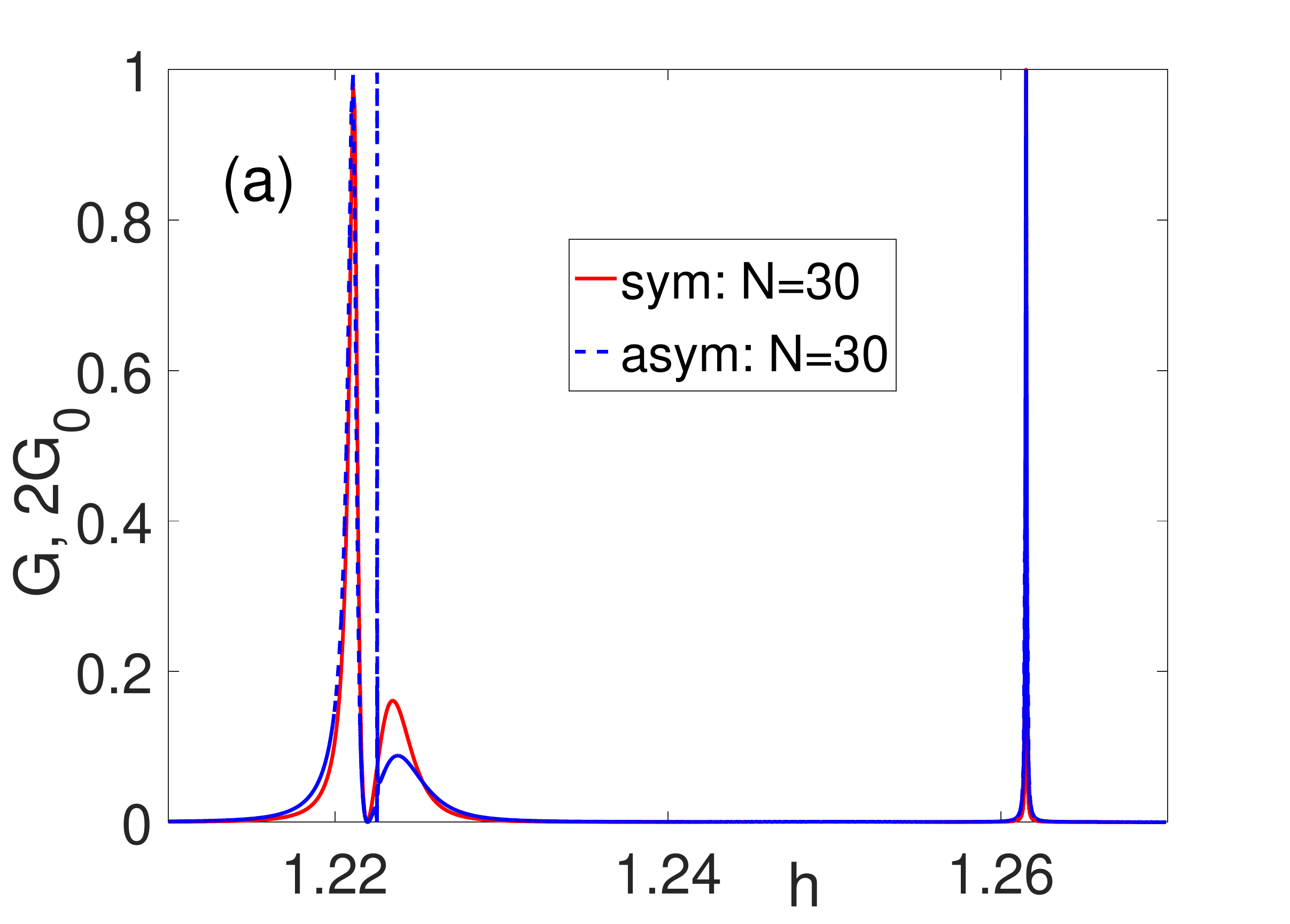}
	\includegraphics[width=0.5\textwidth]{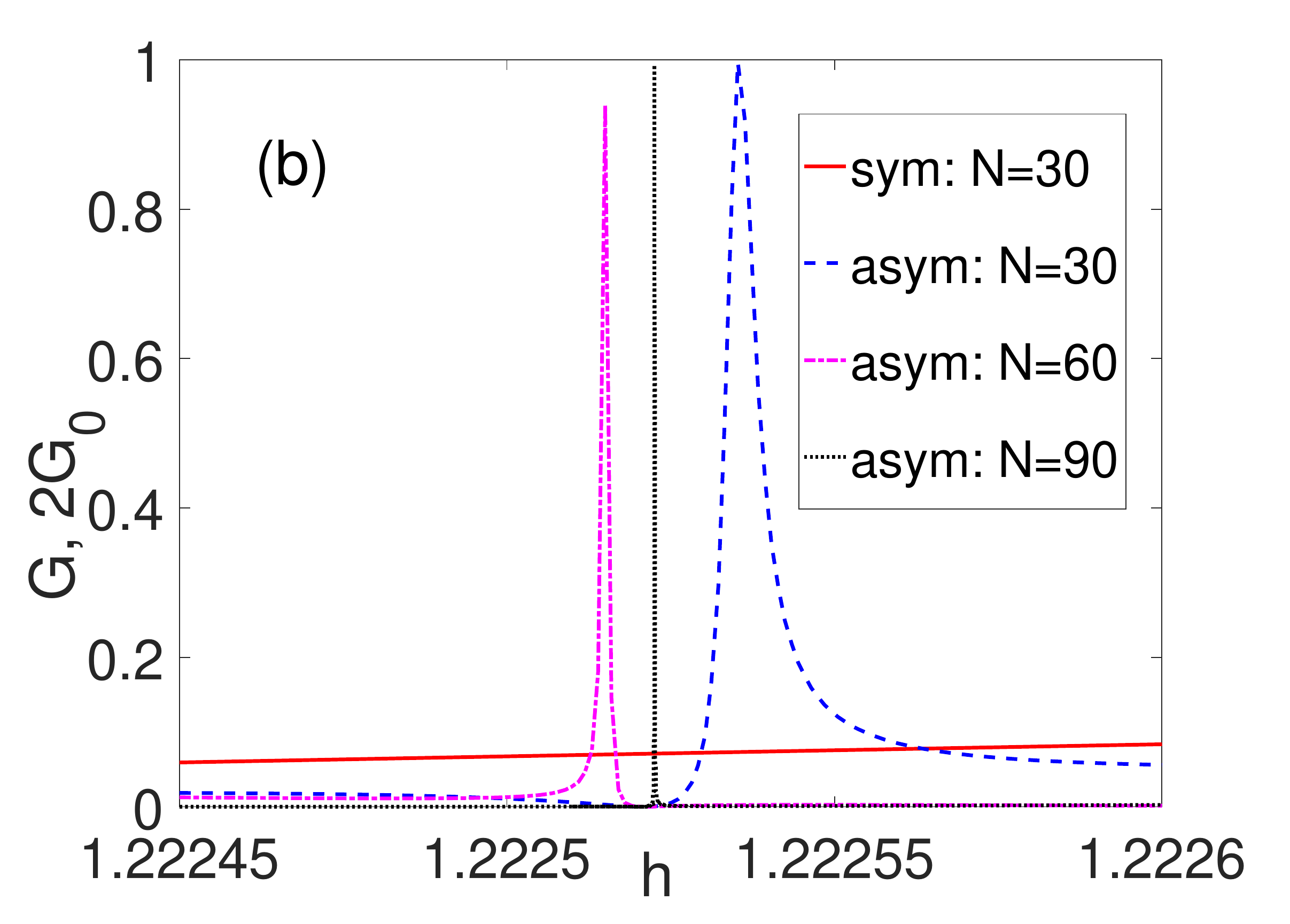}
	\caption{\label{Gh_asym} Fig. \ref{Gh_asym}. (a) Fano resonance arising due to the asymmetry of the tunneling processes in the ring. (b) Collapse of the Fano resonance illustrated in panel (a) with the increase in the degree of MS nonlocality. The values of parameters are $n=20$, $N=30$, $t_{0}=0.1$, $\epsilon_{d}=1$.}
\end{figure}

If the NWs and SW in the ring contain a larger number of sites, $N=30$, $n=20$, respectively, the Dicke effect also takes place, when conditions \eqref{TPineq} are satisfied and $\alpha\neq0$ \cite{valkov-19}. This regime means the implementation of a topologically nontrivial phase in the SW. The solid curve in Fig. \ref{Gh_asym}a denotes a pair of resonances, Fano and Breit-Wigner ones, in the conductance of a symmetric ring ($n=20$, $N=30$) as a function of Zeeman energy. As it has been mentioned above, the properties of the Fano resonance in this case depend on the degree of localization of the MS, which makes it possible to use such device for the detection of these excitations.

Additional features in the electron transport related to the nonlocality of the MS occur in the case of an asymmetric ring. In this case, an additional narrow Fano resonance arises near the wide antiresonance (see the dashed curve in Fig.  \ref{Gh_asym}a). It is important to note that with the increase of the bridge length, the wide antiresonance approaches the narrow Fano peak. In its turn, the latter collapses, which is clearly seen in Fig. \ref{Gh_asym}b, and the BSC appears. In other words, one can speak about the peculiar (topological) blockade of the Fano effect, associated with the asymmetry of the transport processes in the ring, since the corresponding resonance disappears just due to the nonlocality of the low-energy excitation in the SW. 

To get a better insight into the mechanism underlying the collapse of the Fano resonance, it is important to recall that the latter is determined by the BSC arising due to the degeneracy of the eigenstates of a closed system with zero energy. Hence, the vanishing of the Fano resonance could suggest an increase in the multiplicity of the degeneracy of this state if the overlap of the Majorana wave functions becomes negligible. To test this hypothesis, we consider the spinless model of the ring with $n=1$. In this situation, we use the Kitaev chain with an even number of sites in the bridge \cite{kitaev-01}. Then, at  $\epsilon_{d}=\mu=0$ the Hamiltonian of the ring has the form
\begin{eqnarray}\label{HK}
&&H_{D}=\sum\limits_{j=1}^{N-1}\bigl(-ta_{j}^{+}a_{j+1}+\Delta a_{j}^{+}a_{j+1}^{+})-\\
&&\bigl.~~~~~~~~-t_{0}a_{1}^{+}\left(b_{Ln}+b_{Rn}\right)-t_{0}a_{N}^{+}\left(d_{L1}+d_{R1}\right)+h.c.\nonumber
\end{eqnarray}
The diagonalization of Hamiltonian \eqref{HK} leads to the following equation for the excitation spectrum
\begin{eqnarray}\label{CMs}
&&E^{4}\left(E\cdot P_{1}-2t_{0}^{2}\delta_{1}^{N/2-1}\right)\left(E\cdot P_{2}+2t_{0}^{2}\delta_{1}^{N/2-1}\right)\cdot\\
&&~~~~~~~~~\cdot\left(E\cdot P_{3}-2t_{0}^{2}\delta_{2}^{N/2-1}\right)\left(E\cdot P_{4}+2t_{0}^{2}\delta_{2}^{N/2-1}\right)=0,\nonumber
\end{eqnarray}
where $\delta_{1,2}=t \mp \Delta$; $P_{i}$ is the $i$th polynomial of power  $N/2$ for which $P_{2,4}=P_{1,3}\left(E\to-E\right)$ due to the electron?hole symmetry. It follows from \eqref{CMs} that at the specific case of the Kitaev model, $\Delta=\pm t$, where the wave functions of the Majorana fermions do not overlap, the multiplicity of the degeneracy for the zero-energy state increases at $N>2$. This is just the cause of the suppression of the narrow Fano resonance illustrated in Fig. \ref{Gh_asym}b.

\begin{figure}[htbp]
	\includegraphics[width=0.5\textwidth]{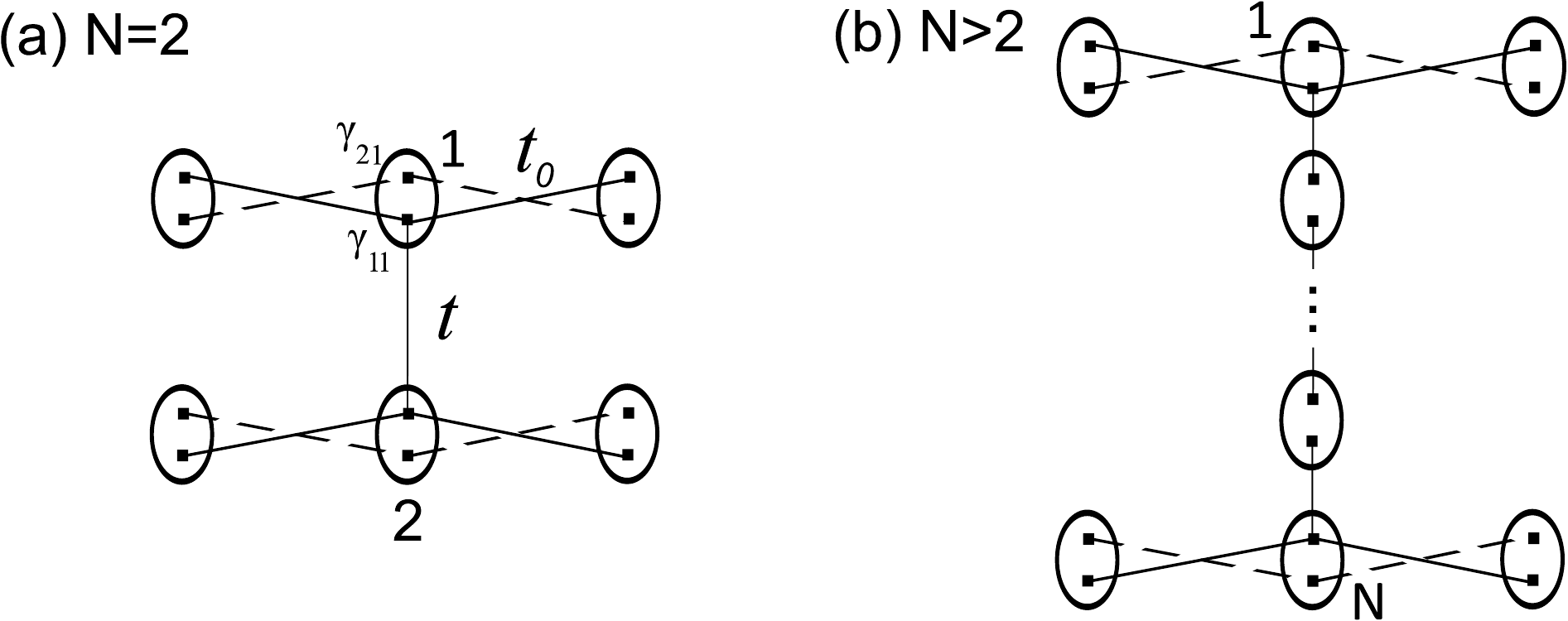}
	\caption{\label{ringMaj} Fig. \ref{ringMaj}. Ring with $n=1$ in the representation of Majorana operators at (a) $N=2$ and (b) $N>2$ for $\Delta=t$.}
\end{figure}

To make the situation clearer, let us turn to the study of our system using the Majorana representation, $a_{j}=\left(\gamma_{1j}+i\gamma_{2j}\right)/2$, where $\gamma_{ij}=\gamma_{ij}^{+}$ ($i=1,2$). In Figs. \ref{ringMaj}a and \ref{ringMaj}b, we schematically present the device in the framework of such description at the specific case of Kitaev model, $\Delta=t$, for $N=2$ and $N>2$, respectively (straight lines denote the interaction between Majorana fermions of different kinds). We can see that in the first case, the upper and lower arms remain connected due to the absence of superconducting pairing in horizontal directions. In the second case, the device is divided into upper and lower identical subsystems. Each of them includes two chains of interacting quasiparticles. The self-energies of a chain with only two bonds in the horizontal direction are $E_{1}=0$ and $E_{2,3}=\pm t_{0}/\sqrt{2}$. If the vertical bond is included (similarly to the Fano-Anderson model), we have $E_{1,2}=0$ and $E_{3,4}=\pm \sqrt{t^2+t_{0}^2/2}$. Thus, it is the formation of the T-shaped structures of Majorana fermions that leads to the suppression of the Fano resonance in the asymmetric ring. Note that the nonlocality of the MS does not depend on the ratio of the tunneling parameters between the subsystems (contacts, NW, and SW), therefore, the effect under discussion has the universal nature and arises in the most general situation typical of experiment, namely, when all these parameters are different. In addition, from Fig. \ref{ringMaj}, it becomes clear that simply at $t=0$, i.e. in the case of two noninteracting arms, the Fano resonance is not suppressed. We should emphasize that in the symmetric case, the described Fano resonance does not arise in principle.

{\bf 4.} In this work, we have analyzed the characteristic features of low-energy quantum transport related to the asymmetry of kinetic processes in the Aharonov-Bohm ring, the arms of which are connected by a SW exhibiting a topologically nontrivial phase. It is found that the Fano resonance arising due to such symmetry breaking collapses with an increase in the bridge length or, in other words, when the overlap of the Majorana wave functions becomes negligible. To get a better insight into this effect, we have considered the model of a spinless ring, in which the Kitaev chain acts as a SW. The analytical calculation of the spectrum of such system reveals an increase in the multiplicity of the degeneracy of the zero-energy state at specific case of the Kitaev model at $N>2$ due to the formation of the T-shaped chains of Majorana fermions. This is direct consequence of the nonlocality of the MS.

\begin{acknowledgments}
We are grateful to V.V. Valkov and A.D. Fedoseev for stimulating discussions. The work was supported by the Presidium of the Russian Academy of Sciences, Program of basic research no. 32 "Nanostructures: physics, chemistry, biology, and fundamentals of technologies", by the Russian Foundation for Basic Research (project nos. 19-02-00348, 20-32-70059, and 20-02-00015), by the Government of the Krasnoyarsk Territory, and by the Krasnoyarsk Science Foundation, project no. 19-42-240011 "Coulomb interactions in the problem of Majorana modes in low-dimensional systems with nontrivial topology". S.V. Aksenov acknowledges the support from the Council of the President of the Russian Federation for Support of Young Russian Scientists and Leading Scientific Schools, grant no. MK-3722.2018.2. M.Yu. Kagan acknowledges the support from the Program of Basic Research of the National Research University Higher School of Economics.
\end{acknowledgments}

\end{document}